\documentclass[12pt]{article}
\usepackage{amssymb}
\usepackage{epsfig}
\usepackage{amssymb,amsmath}
\usepackage{slashed}
\usepackage{multirow}

\newcommand{\bea}{\begin{eqnarray}}
\newcommand{\eea}{\end{eqnarray}}
\newcommand{\beq}{\begin{equation}}
\newcommand{\eeq}{\end{equation}}

 \makeatletter
\def\alt{\mathrel{\mathpalette\gl@align<}}
\def\agt{\mathrel{\mathpalette\gl@align>}}
\def\gl@align#1#2{\lower.6ex\vbox{\baselineskip\z@skip\lineskip\z@
\ialign{$\m@th#1\hfil##\hfil$\crcr#2\crcr\sim\crcr}}} \makeatother

\begin{document}

%
\vspace*{1.0cm}

\begin{center}
\baselineskip 20pt {\Large\bf  Supersymmetric Hybrid Inflation Redux} \vspace{1cm}

{\large Mansoor Ur Rehman, Qaisar Shafi, Joshua R. Wickman} \vspace{.5cm}

{\baselineskip 20pt \it 
Bartol Research Institute, Department of Physics and Astronomy, \\
University of Delaware, Newark, Delaware 19716, USA \\
}
August 2009 \\

\vspace{.5cm}

\end{center}

\begin{abstract}
We discuss the important role played during inflation by one of the soft supersymmetry breaking terms in the inflationary potential of supersymmetric hybrid inflation models. With minimal K\"ahler potential, the inclusion of this term allows the prediction for the scalar spectral index to agree with the value $n_s = 0.963^{+0.014}_{-0.015}$ found by WMAP5. In the absence of this soft term, and by taking into account only radiative and supergravity corrections, it is well known that $n_s \geq 0.985$. This same soft term has previously been shown to play a key role in resolving the MSSM $\mu$ problem. The tensor to scalar ratio $r$ is quite small in these models, taking on values $r \leq 10^{-5}$ in the WMAP5 $2\sigma$ range of $n_s$.

\end{abstract}

\date{}

Supersymmetric (SUSY) hybrid inflation models \cite{Dvali:1994ms,Copeland:1994vg,Lazarides:2001zd,Linde:1993cn}, incorporating low ($\sim$ TeV) scale supersymmetry and minimal ($N=1$) supergravity (SUGRA) \cite{Chamseddine:1982jx}, have attracted a great deal of attention because of their intimate connections to mainstream particle physics \cite{Senoguz:2003zw}. Inflation in these models is typically realized within the framework of a supersymmetric gauge theory based on a gauge group $G$, which is spontaneously broken after inflation is over to its subgroup $H$ (regular inflation), or during the inflationary phase (shifted inflation \cite{Jeannerot:2000sv}). There are several important features of supersymmetric hybrid inflation which are worth noting here. First, the symmetry breaking scale $M$ is determined by the consistency of the inflationary scenario
and turns out to be comparable to $M_{GUT} \sim 10^{16}$ GeV. Roughly speaking, the CMB anisotropy $\delta T/T$ is proportional to $( M/M_P)^2$, which explains why $M$ is of order $M_{GUT}$ \cite{Dvali:1994ms}. Second,  for a wide range of parameters, the inflaton field takes values that are not much larger than $M_{GUT}$. Thus, in contrast to chaotic inflation in which the inflaton acquires trans-Planckian values, the supergravity corrections in supersymmetric hybrid inflation models are adequately suppressed  \cite{Linde:1997sj,Senoguz:2004vu}. Third, it was shown in Ref.~\cite{Senoguz:2004vu} that one of the soft supersymmetry breaking terms generated in minimal supergravity, can play an important role during inflation. This term had earlier been shown to play a key role in the resolution of the MSSM $\mu$ problem \cite{Dvali:1997uq}. Finally, if $G$ is identified with $\left[ SU(3)\times SU(2) \times U(1) \right] \times U(1)_{B-L}$ or $SO(10)$ for example, the observed baryon asymmetry can be generated in these models via non-thermal leptogenesis \cite{Lazarides:1996dv}.
The main purpose of this letter is to carefully investigate the role during inflation of the soft supersymmetry breaking term. We find that its inclusion in the inflationary potential gives rise to a scalar spectral index which is smaller than $0.985$, the previous lower bound obtained in these models, and therefore in better agreement with the central value of $0.963$  estimated by the WMAP 5 yr analysis (WMAP5) \cite{Komatsu:2008hk}. Note that this result is achieved by employing the canonical (minimal) K\"ahler potential. For a discussion involving a non-minimal K\"ahler potential, see Refs.~\cite{BasteroGil:2006cm,urRehman:2006hu}. 

SUSY hybrid inflation is defined by 
the superpotential $W$ \cite{Copeland:1994vg,Dvali:1994ms}
\begin{equation} 
W=\kappa \hat{S}(\hat{\Phi} \hat{\overline{\Phi }}-M^{2})\,,
\label{superpot}
\end{equation}
where $\hat{S}$ is a gauge singlet superfield and $\hat{\Phi}$, $\ \hat{\overline{\Phi }}$ are 
a conjugate pair of superfields transforming as
nontrivial representations of some gauge group $G$. As a simple example, $G$ can be the
standard model gauge group supplemented by a gauged $U(1)_{B-L}$, which
requires, for anomaly cancellations, the presence of three right handed
neutrinos. $W$ has a $U(1)_R$ `$R$-symmetry' such that $W\rightarrow e^{i\alpha}W$, $\hat{S}\rightarrow e^{i\alpha}\hat{S}$, $\hat{\Phi}\hat{\overline{\Phi}}\rightarrow \hat{\Phi}\hat{\overline{\Phi}}$
and it can readily be seen that Eq.~(\ref{superpot}) is the most general renormalizable superpotential  consistent with both $R$-symmetry and $G$-invariance.

The SUGRA scalar potential is given by
\begin{equation}
V_{F}=e^{K/m_{P}^{2}}\left(
K_{ij}^{-1}D_{z_{i}}WD_{z^{*}_j}W^{*}-3m_{P}^{-2}\left| W\right| ^{2}\right),
\label{VF}
\end{equation}
with $z_{i}$ being the bosonic components of the superfields $z%
_{i}\in \{\Phi , \overline{\Phi }, S,\cdots\}$ and where we have defined
\begin{eqnarray*}
D_{z_{i}}W &\equiv &\frac{\partial W}{\partial z_{i}}+m_{P}^{-2}\frac{%
\partial K}{\partial z_{i}}W, \,\,\,\,\,\, K_{ij} \equiv \frac{\partial ^{2}K}{\partial z_{i}\partial z_{j}^{*}},
\end{eqnarray*}
$D_{z_{i}^{*}}W^{*}=\left( D_{z_{i}}W\right) ^{*}$ and $m_{P}=M_P/\sqrt{8\pi} \simeq 2.4\times 10^{18}$ GeV is the reduced Planck mass. The minimal K\"ahler potential can be expanded as
\begin{equation}
K=  |\hat{S}|^{2}+ |\hat{\Phi}|^{2} + |\hat{\overline{\Phi}}|^{2}.  \label{kahler}
\end{equation}
In the D-flat direction $|\Phi|=|\overline{\Phi}|$, 
and using Eqs.~(\ref{superpot},~\ref{kahler}) in Eq.~(\ref{VF}), the tree level global SUSY potential is given by

\begin{equation}
V_{F}= \kappa^2\,(M^2 - \vert \Phi\vert^2)^2 + 2\kappa^2 \vert S \vert^2 \vert \Phi \vert^2 .
\label{VF2}
\end{equation}
Assuming suitable initial conditions, the fields get trapped in the inflationary
valley of local minima at $\left| S\right| >S_{c}=M$ and $\left| \Phi
\right| =\left| \overline{\Phi }\right| =0$, where $G$ is unbroken. The
potential is dominated by the constant term $V_{0}=\kappa ^{2}M^{4}$, thus SUSY is broken during inflation. 
Inflation ends when the inflaton drops below its
critical value $S_{c}=M$ and the fields roll towards the global SUSY minimum of
the potential $\left| S\right| =0$ and $\left| \Phi \right| =\left| 
\overline{\Phi }\right| =M$. 
Taking into account leading order SUGRA corrections,
as well as radiative corrections \cite{Dvali:1994ms} and soft SUSY breaking terms, the potential along the inflationary trajectory ($|\Phi| = |\overline{\Phi}| = 0$) is of the form 
\begin{eqnarray}
V &\approx& \kappa ^{2}M^{4}\left( 1 + \left( \frac{M}{m_{P}}\right) ^{4}\frac{x^{4}}{2}+\frac{%
\kappa ^{2}\mathcal{N}}{8\pi ^{2}}F(x) + a\left(\frac{m_{3/2}\,x}{\kappa\,M}\right) + \left( \frac{m_{3/2}\,x}{\kappa\,M}\right)^2\right) ,
\label{scalarpot}
\end{eqnarray}
where 
\begin{equation}
F(x)=\frac{1}{4}\left( \left( x^{4}+1\right) \ln \frac{\left( x^{4}-1\right) 
}{x^{4}}+2x^{2}\ln \frac{x^{2}+1}{x^{2}-1}+2\ln \frac{\kappa ^{2}M^{2}x^{2}}{%
Q^{2}}-3\right),
\end{equation}
and 
\begin{equation}
a = 2\left| 2-A\right| \cos [\arg S+\arg (2-A)].
\label{a}
\end{equation}
Here $\mathcal{N}$ is the dimensionality of the representation of the fields 
$\Phi$ and $\overline{\Phi },$ $Q$ the renormalization scale and $x=\left|
S\right| /M.$ In Eq.~(\ref{scalarpot}), the $F(x)$-term represents the radiative corrections, while the
last two terms are the soft SUSY breaking linear and mass-squared terms, respectively.  The quantity $A-2$ represents the coefficient of the linear soft term~\cite{Nilles:1983ge}, and the magnitude of $A$ is expected to be of order unity or so.  The form of these soft terms comes from gravity-mediated SUSY breaking~\cite{Dvali:1997uq}, which we employ throughout. Relative to the linear soft term, the soft mass term is subdominant for $|a| \gtrsim 10^{-6}$.  [Note that the soft linear term induces a VEV proportional to $m_{3/2}$ for $S$, which has been exploited in Ref.~\cite{Dvali:1997uq} to resolve the MSSM $\mu$ problem.]
By virtue of the minimal K\"ahler potential, an $|S|^2$ term is not induced by SUGRA corrections, as this contribution is canceled by an identical term arising from the superpotential.  As a result, the $\eta$ problem is under control in these `minimal' models.

In our numerical calculations, we will take $m_{3/2} = 1$ TeV and $\mathcal{N}=1$ for simplicity. As pointed out in Ref. \cite{Senoguz:2004vu}, $\arg S$ can vary significantly for 
$\kappa \sim 10^{-4}$, yet it is always possible to suppress this variation via the choice of initial conditions, as explicitly shown in Ref.~\cite{urRehman:2006hu}. For $\kappa$ values much different from $10^{-4}$, the variation of $\arg S$ is negligibly small.  Therefore, we will assume that $\arg S$ is constant during inflation, and will study only the specific cases $a = 0$ and $a = -1$. For an analysis involving $a>0$, see Refs.~\cite{Senoguz:2004vu,Jeannerot:2005mc}. 

For $a \geq 0$, the potential in Eq.~(\ref{scalarpot}) monotonically increases with $x$.
In contrast, for $a < 0$, the contribution from this term can dominate in some region of parameter space for inflation beginning around $x \sim 1$.  In this case, the potential develops a local maximum, giving rise to `hilltop inflation' beginning near this maximum \cite{Boubekeur:2005zm}. 
Recent treatments have shown that hilltop solutions exist in these models if a non-minimal K\"ahler potential is used \cite{Pallis:2009pq}; here, we are able obtain such solutions by employing the minimal K\"ahler potential. 
It is interesting to note that hilltop solutions are also generated in models of non-SUSY hybrid inflation which include fermion-dominated radiative corrections \cite{Rehman:2009wv}.

The number of $e$-folds after the comoving scale $l$ has crossed the horizon
is given by
\begin{equation}
N_{l}=2\left( \frac{M}{m_{P}}\right) ^{2}\int_{x_e}^{x_{l}}\left( \frac{V}{%
\partial _{x}V}\right) dx , \label{Nl}
\end{equation}
where $|S_{l}|=x_{l}\,M$  is the field value at the comoving scale $l$, and $x_e$ denotes the value of $x$ at the end of inflation.  If the slow-roll approximation holds, inflation ends via a waterfall induced at $x_e=1$, although in practice the slow-roll conditions are typically violated at values of $x_e$ slightly larger than unity.
During inflation, the comoving scale corresponding to $k_{0}=2\pi/l_0=0.002$~Mpc$^{-1}$ exits the horizon
such that the number of $e$-folds is approximately given by
\begin{equation}
N_{0} \simeq 53+\frac{1}{3}\ln \left( \frac{T_{r}}{10^{9}~{\rm GeV}}\right) +\frac{2}{3}%
\ln \left( \frac{V(x_0)^{1/4}}{10^{15}~{\rm GeV}}\right),   \label{N0}
\end{equation}
where $T_{r}$ is the reheat temperature, and the subscript `0' 
indicates that the values are taken at $k_{0}$.

After the end of inflation, the fields fall toward the SUSY vacuum and undergo damped oscillations about it.  The coupled $S$,~$\Phi$,~$\overline{\Phi}$ system then decays into right-handed neutrinos and sneutrinos, which in turn give rise to the observed baryon asymmetry via non-thermal leptogenesis \cite{Lazarides:1996dv,Lazarides:1991wu}. In this case, the reheat temperature is well approximated by \cite{Senoguz:2004vu}
\begin{equation}\label{trmin}
T_r\gtrsim1.6\times10^{7}{\rm\ GeV}\left(\frac{10^{16}{\rm\
GeV}}{M}\right)^{1/2}\left(\frac{m_{\rm inf}}{10^{11}{\rm\ GeV}}\right)^{3/4}
\left(\frac{0.05\rm{\ eV}}{m_{\nu3}}\right)^{1/2}\,, 
\end{equation}
where $m_{\rm inf}=\sqrt{2}\kappa M$ is the mass of the inflaton, and we will take the mass of the heaviest light neutrino to be $m_{\nu3}\simeq 0.05$~eV.

The amplitude of the curvature perturbation is given by
\begin{equation}
\Delta_{\mathcal{R}}=\frac{M}{\sqrt{6}\,\pi \,m_{P}^{3}}\left( \frac{V^{3/2}}{|\partial
_{x_{0}}V|}\right)  , \label{curv}
\end{equation}
where $\Delta_{\mathcal{R}}=4.91\times 10^{-5}$ is the WMAP5 normalization at $k_{0}$ \cite{Komatsu:2008hk}.
For $\mathcal{N}=1$, corresponding say to the breaking of a U(1) gauge symmetry, cosmic strings are produced upon symmetry breaking.  The contribution of cosmic strings to the curvature perturbation goes as the string tension, $(\delta T/T)_{cs} \propto G\mu \sim (M/m_P)^2$~\cite{Jeannerot:2005mc}.  As we will see, the values of $M$ obtained in these models are roughly $\sim 1.1\times 10^{15}$~GeV for $n_s\sim 0.963$, and so the contribution from cosmic strings is subdominant.  In our calculations, we will suppress this relatively small ($\lesssim$ a few percent) contribution from cosmic strings, and consider the curvature perturbation to arise from primordial inflation to within the uncertainty in its measured value.

\begin{figure}[t]
\centering \includegraphics[angle=0, width=8cm]{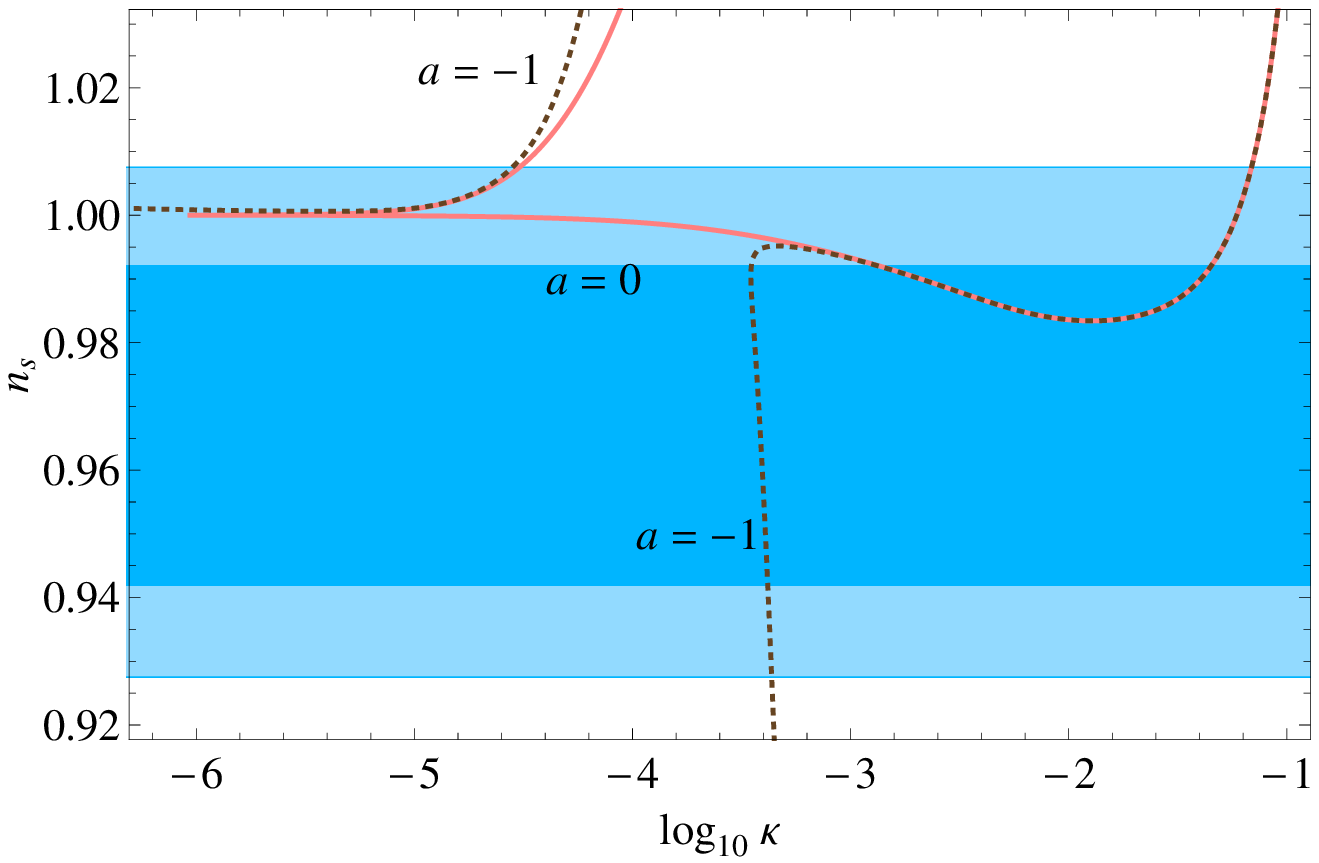}
\centering \includegraphics[angle=0, width=8cm]{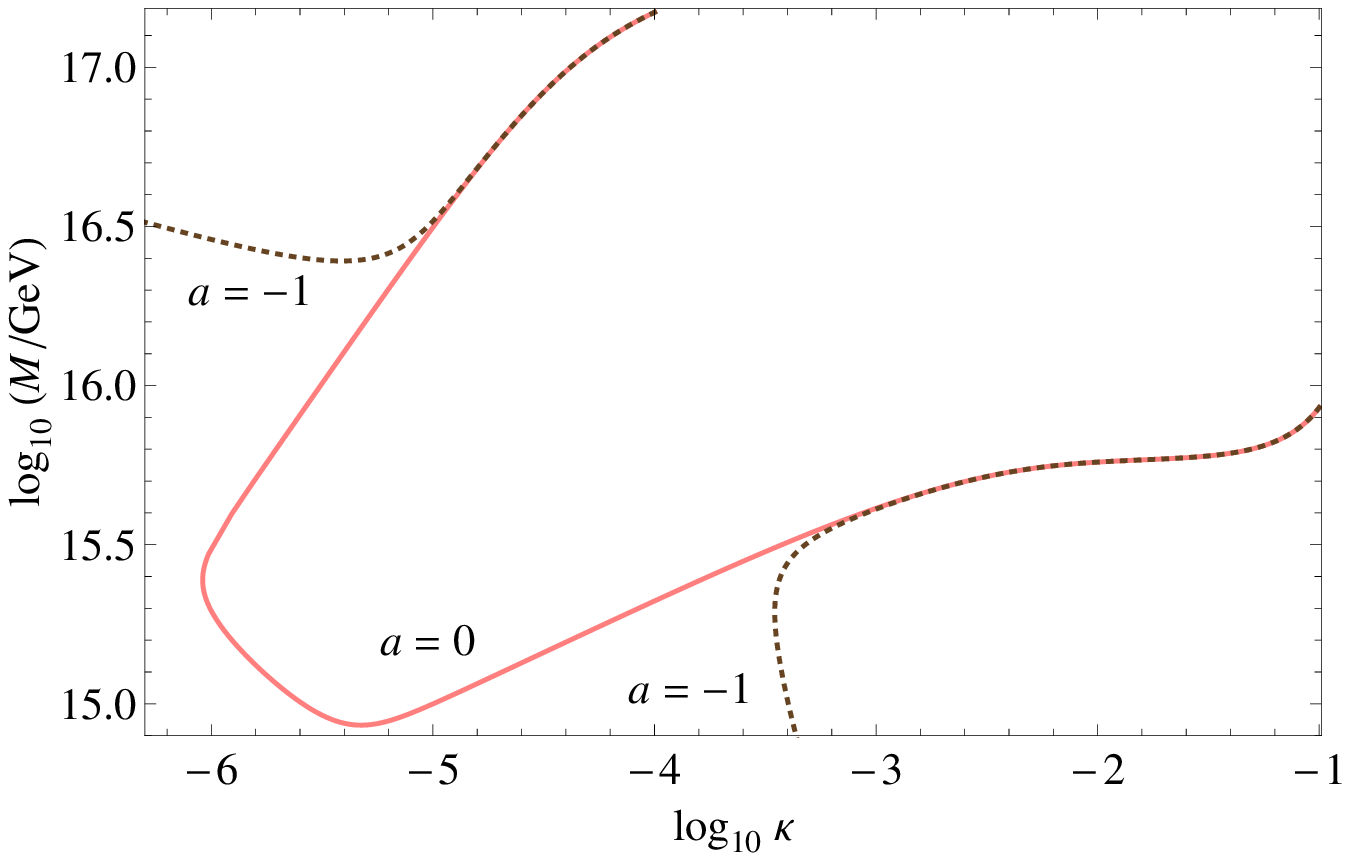}
\caption{$n_s$ and $\log_{10}(M/\text{GeV})$ vs. $\log_{10}\kappa$,
for $m_{3/2} = 1$ TeV, $a = 0$ and $a = -1$, with $a$ given by Eq.~(\ref{a}).  
The left panel includes the WMAP5 68\% and 95\% confidence level contours for the value of the spectral index $n_s$ \cite{Komatsu:2008hk}.}
\label{nsMk}
\end{figure}

The usual slow-roll parameters may be defined as
\begin{equation}
\epsilon =\frac{m_{P}^{2}}{4\,M^2}\left( \frac{\partial_x V }{V}\right) ^{2},\,\,
\eta =\frac{m_{P}^{2}}{2\,M^2}\left( \frac{\partial_x^2 V }{V}\right), \,\,
\xi ^{2} = \frac{m_{P}^{4}}{4\,M^4}\left( \frac{\partial_x V~ \partial_x^3 V }{V^{2}}\right).
\end{equation}
In the slow-roll approximation (i.e. $\epsilon,|\eta|,\xi^2\ll 1$), the spectral index $n_{s}$ is given (to leading order) by 
\begin{eqnarray}
n_{s} &\simeq& 1 -6\epsilon +2\eta \,\, \simeq \,\, 1 + 6\left( \frac{M}{m_{P}}\right)
^{2}x_{0}^{2}+\left( \frac{m_{P}}{M}\right) ^{2}\left[ \frac{\kappa ^{2}%
\mathcal{N}}{8\pi ^{2}} \partial _{x_0}^{2}F(x_0)+ 2 \left(\frac{m_{3/2}}{\kappa\,M}\right)^2 \right]\,.\label{ns}
\end{eqnarray}
Using Eqs.~(\ref{scalarpot})--(\ref{curv}), we have calculated $n_{s}$ and $M$ as a function of $\kappa $ for $a = -1$ and $a = 0$, as shown in Fig.~\ref{nsMk}.  (In these numerical calculations, we have used the next-to-leading order expressions in the slow-roll approximation for added precision.)
Similar to the $a>0$ case, we obtain two branches of solutions for $a \leq 0$, one appearing for low values and another for high values of $M$. For $a=-1$ these two branches are disconnected, whereas for $a=0$, they merge in the region where the soft SUSY mass term is important.
We find that a red spectral index in agreement with the WMAP5 central value, $n_s=0.963$, can be generated for negative values of $a$ with magnitude $|a| \gtrsim 10^{-6}$.  For smaller $|a|$, a solution coinciding with $a=0$ in the $n_s$-$\kappa$ plane is abruptly reached, and a red tilted spectrum is not exhibited for $\kappa \lesssim 10^{-5}$.

The slow-roll parameter $\epsilon$ is quite small in these models, and its contribution to $(n_s-1)$ is masked by the $\eta$ term.  Thus we have suppressed this term in the right-hand side of Eq.~(\ref{ns}).  The form of this equation suggests that $(n_s-1)$ receives a positive contribution from all terms except the radiative correction term, which turns out to be negative.  However, the role played by the $a$-term is not readily apparent from this approximate equation; this term contributes most directly through the $\Delta_{\mathcal{R}}$ constraint (encoded via $x_0$), and it can be seen from Fig.~\ref{nsMk} that its inclusion can lead to qualitatively different results.  We now turn to a more quantitative analysis of the role of the $a$-term.

From Eq.~(\ref{scalarpot}), we see that the SUGRA term dominates the potential if $x\gg 1$.  In other words, the limit $x_0\sim 1$ is a good approximation in regions where the terms in the potential may compete with one another.  In this limit, Eq.~(\ref{curv}) becomes
\begin{equation}
\Delta_{\mathcal{R}} \simeq \frac{\kappa^2}{2\sqrt{6}\pi} \left( \frac{M}{m_P} \right)^4 \left[ \kappa\left(\frac{M}{m_P}\right)^5 + \frac{\kappa^3 \mathcal{N} \ln 2}{8\pi^2} \left(\frac{M}{m_P}\right) + \frac{a m_{3/2}}{2 m_P} + \frac{m_{3/2}^2}{\kappa M m_P} \right]^{-1}.\label{curvlim}
\end{equation}
By fixing $\Delta_{\mathcal{R}}$ at the WMAP5 value and differentiating, it is possible to find limiting values of $\kappa$ and $M$ with respect to one another in regions where two of the terms in Eq.~(\ref{curvlim}) compete.  As seen in Fig.~\ref{nsMk}, each case for $a$ will lead to a limiting for each of $\kappa$ and $M$ (on separate branches in the $a<0$ case).

The limiting values of $\kappa$ may be obtained by differentiating Eq.~(\ref{curvlim}) with respect to $M$ and setting $\partial\kappa/\partial M=0$.  For $a=0$, $\kappa$ is small in the limiting region, and the contribution from radiative corrections is negligible.  The interaction of the SUGRA and SUSY mass terms then leads to
\begin{equation}
\kappa_{a=0} \gtrsim \left[ \frac{2^{15} 3^9 \pi^6}{5^5} \left( \frac{m_{3/2}}{m_P} \right)^2 \Delta_{\mathcal{R}}^6 \right]^{1/8} \approx 9.1\times 10^{-7}, \,\,\,\,\, \text{with } M \approx 2.4\times 10^{15} \text{ GeV}.
\end{equation}
In the $a=-1$ case, the SUSY mass term is negligible as compared to the $a$-term, and can be neglected throughout.  In the limiting region of $\kappa$, the SUGRA term is subdominant, and the competition of the radiative correction and $a$-terms yields
\begin{equation}
\kappa_{a=-1} \gtrsim \left[ \frac{2^{31/2} \pi^7}{3^{7/2} (\mathcal{N}\ln 2)^4 \Delta_{\mathcal{R}}} \left( \frac{|a|m_{3/2}}{m_P} \right)^3  \right]^{1/10} \approx 3.4\times 10^{-4}, \,\,\,\,\, \text{with } M \approx 2.0\times 10^{15} \text{ GeV}.
\label{klima}
\end{equation}
Similarly, the limiting values of $M$ are obtained by taking $\partial M/\partial \kappa=0$.  For $a=0$, $\kappa$ is larger than its limit and $M$ is at its smallest value, resulting in suppression of the SUGRA term.  The interplay of the radiative correction and SUSY mass terms then gives
\begin{equation}
M_{a=0} \gtrsim m_P \left[ \left( \frac{2^5}{3} \right)^{1/2} \frac{\Delta_{\mathcal{R}}^2}{\pi} (\mathcal{N}\ln 2)^{3/2} \frac{m_{3/2}}{m_P} \right]^{1/7}
\approx 8.4\times 10^{14} \text{ GeV}, \,\,\,\,\, \text{with } \kappa \approx 4.7\times 10^{-6}.
\end{equation}
Finally, the limiting value of $M$ appears only on the upper branch of results in the $a=-1$ case, where large $M$ and small $\kappa$ values lead to SUGRA dominance over the radiative correction term.  The interplay of the SUGRA and $a$-terms leads to
\begin{equation}
M_{a=-1} \gtrsim m_P \left[ \frac{1}{\sqrt{6}\pi \Delta_{\mathcal{R}}} \frac{|a| m_{3/2}}{m_P} \right]^{1/6}
\approx 2.6\times 10^{16} \text{ GeV}, \,\,\,\,\, \text{with } \kappa \approx 6.0\times 10^{-6}.
\label{Mlima}
\end{equation}
(Note that while the numerical values in Eqs. (\ref{klima}) and (\ref{Mlima}) are given for $a=-1$, we find numerically that the analytical expressions hold for $a \lesssim -10^{-6}$, where the soft SUSY mass term begins to 
play a significant role.)
  For a fixed value of $\kappa$ ($M$) larger than its limit, there exist two values of $M$ ($\kappa$) which satisfy the $\Delta_{\mathcal{R}}$ constraint.

\begin{figure}[t]
\centering \includegraphics[angle=0, width=8cm]{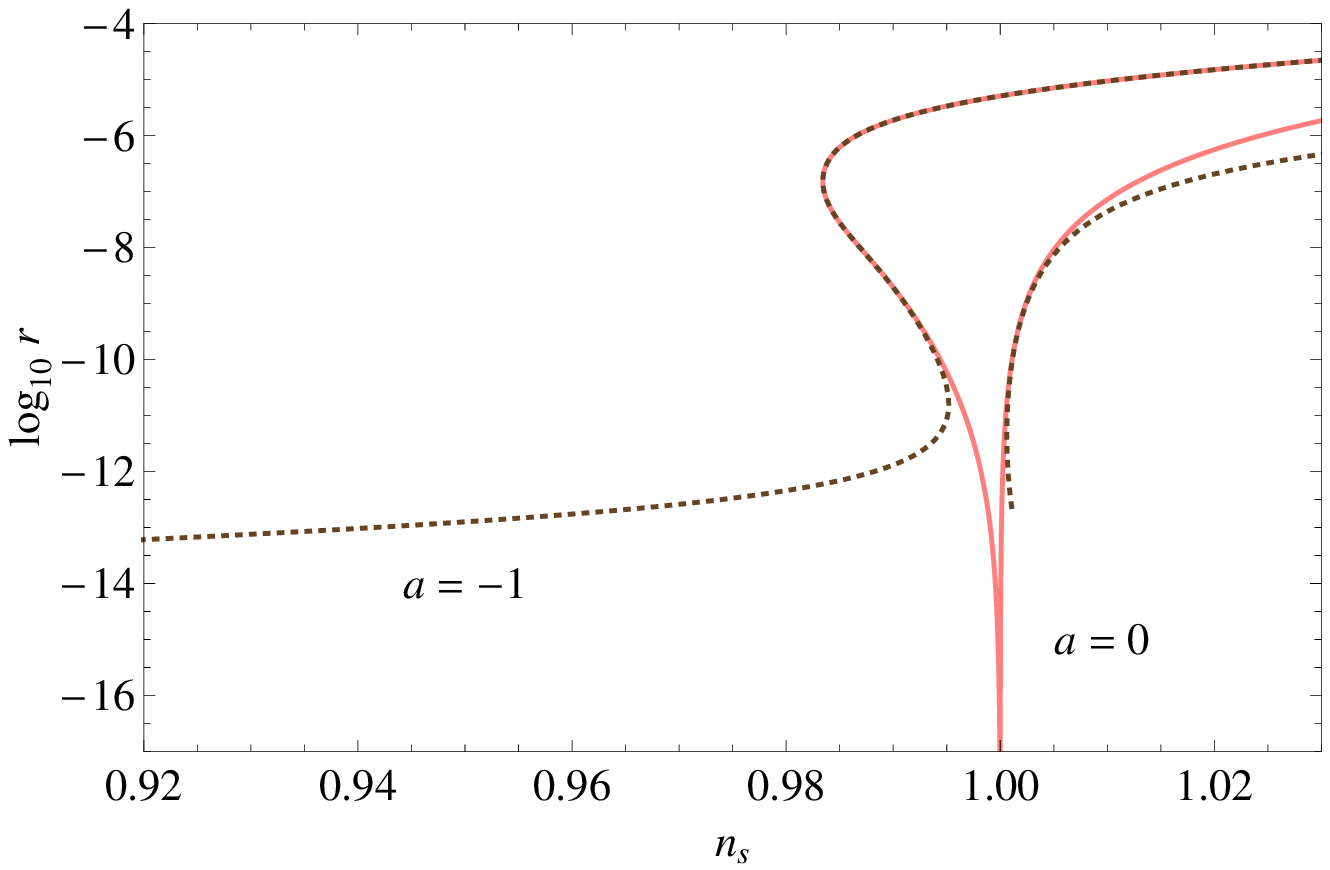}
\centering \includegraphics[angle=0, width=8cm]{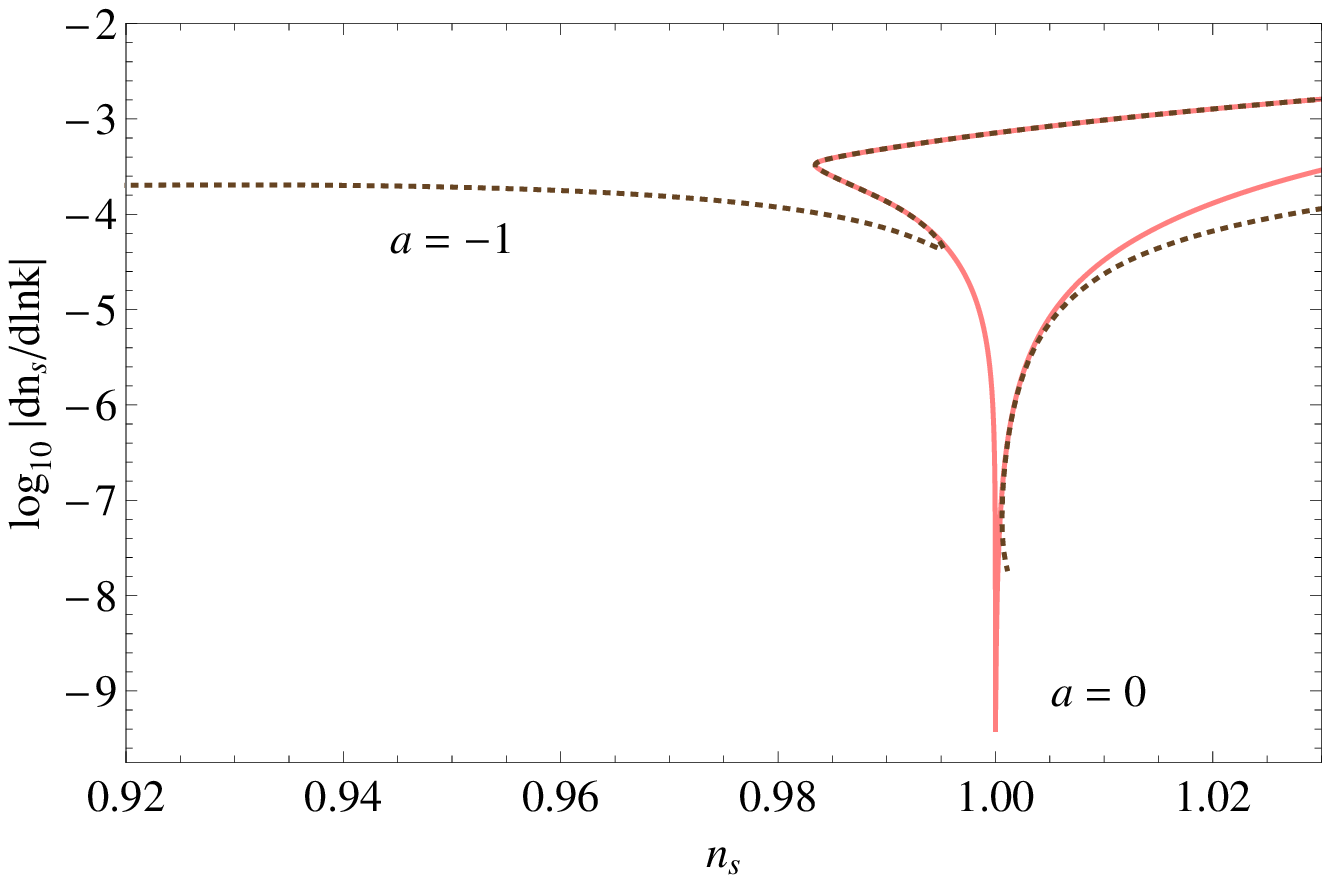}
\caption{$\log_{10}r$ and $\log_{10}|dn_s/d\ln k|$ vs. $n_s$, 
for $m_{3/2} = 1$ TeV, $a = 0$ and $a = -1$, with $a$ given by Eq.~(\ref{a}).} \label{rns}
\end{figure}

The tensor to scalar ratio $r$ is an important cosmological parameter, as it describes the (squared) amplitude of primordial gravitational waves and gives information on the energy scale of inflation.  To leading order in slow-roll, this quantity is given by
\begin{equation}
r \simeq 16 \epsilon \simeq 4 \left( \frac{m_P}{M} \right)^2 \left[ 2 \left( \frac{M}{m_P} \right)^4 x_0^3 + \frac{\kappa^2 \mathcal{N}}{8\pi^2} \partial _{x_0}F(x_0) + \frac{a m_{3/2}}{\kappa M} + 2 \left(\frac{m_{3/2}}{\kappa M}\right)^2 x_0 \right]^2,
\label{r}
\end{equation}
where the first derivative of $F(x)$ is always positive during inflation.  An accurate measurement of $r$ can be quite powerful in discriminating between various classes of inflation models, and even between qualitatively similar models (see, for instance, Ref.~\cite{Rehman:2008qs}).

Experimental constraints on $r$ set by WMAP have become increasingly restrictive, and these bounds will become even more precise with measurements to be taken with the PLANCK satellite.  In SUSY inflation models, $r$ is typically quite small.  In the models which we currently consider, $r$ takes on values $r\lesssim 10^{-5}$ in the region of $n_s$ favored by WMAP5 (see Fig.~\ref{rns}).  

Precision measurements to be performed by PLANCK may also place greater constraints on the running of the spectral index $dn_s/d\ln k$, which we find to be favored at a value of $|dn_s/d\ln k| \sim 10^{-4}$.  This quantity is nonzero only for next-to-leading order in slow-roll
\begin{equation}
\frac{dn_s}{d\ln k} \simeq 16 \epsilon \eta - 24 \epsilon^2 - 2 \xi^2 .
\label{alpha}
\end{equation}
Even so, Fig.~\ref{rns} shows that $|dn_s/d\ln k|$ turns out to be much larger than $r$, particularly in the vicinity of the WMAP5 central value of $n_s$.  This can be understood first by noting that $|\eta| \gg \epsilon$.  Moreover, it can be shown that $\xi^2 \propto \sqrt{\epsilon}$, which is much larger than $\epsilon$ in the region of interest.

\begin{figure}[t]
\centering \includegraphics[angle=0, width=8cm]{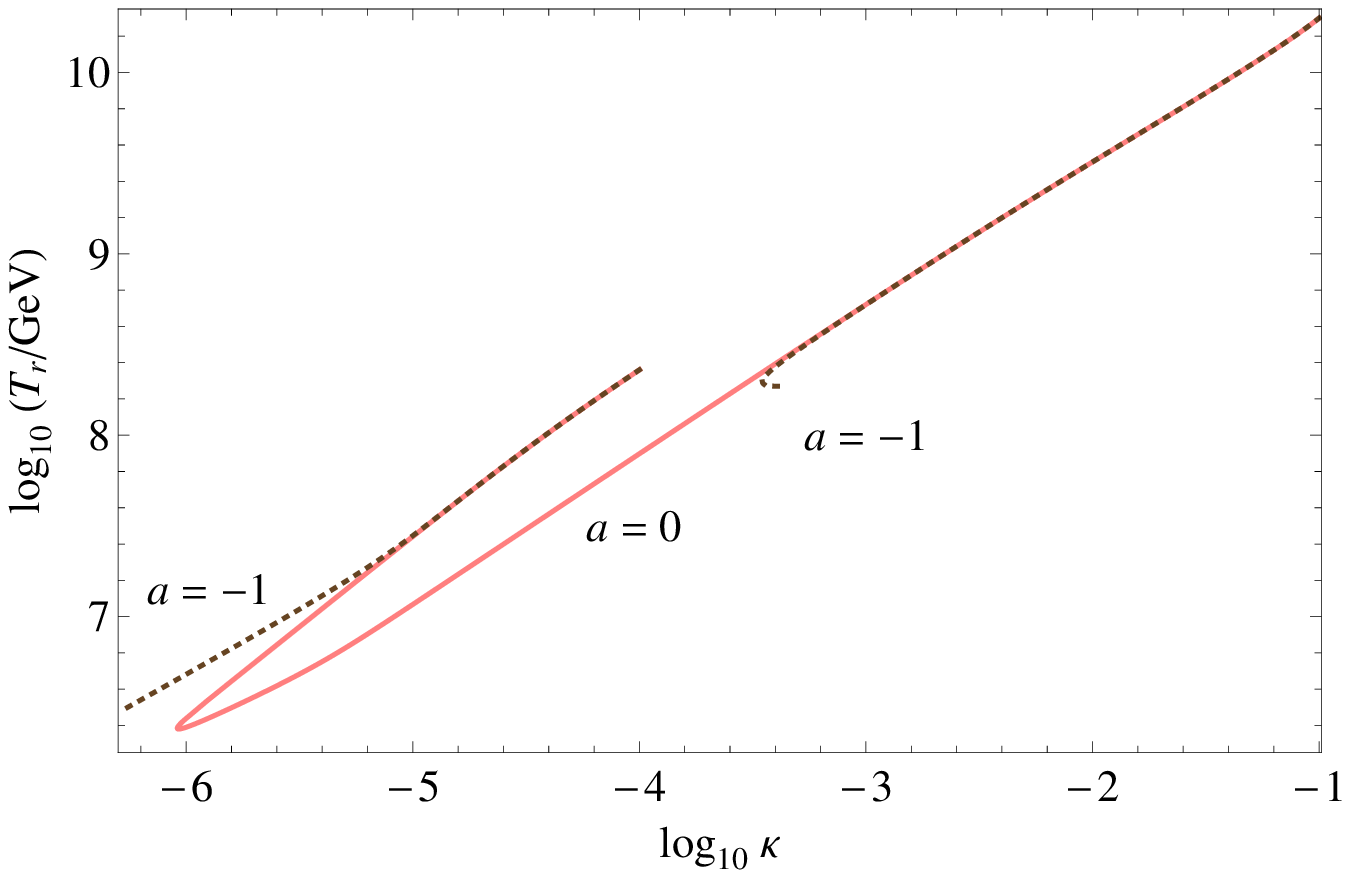}
\centering \includegraphics[angle=0, width=8cm]{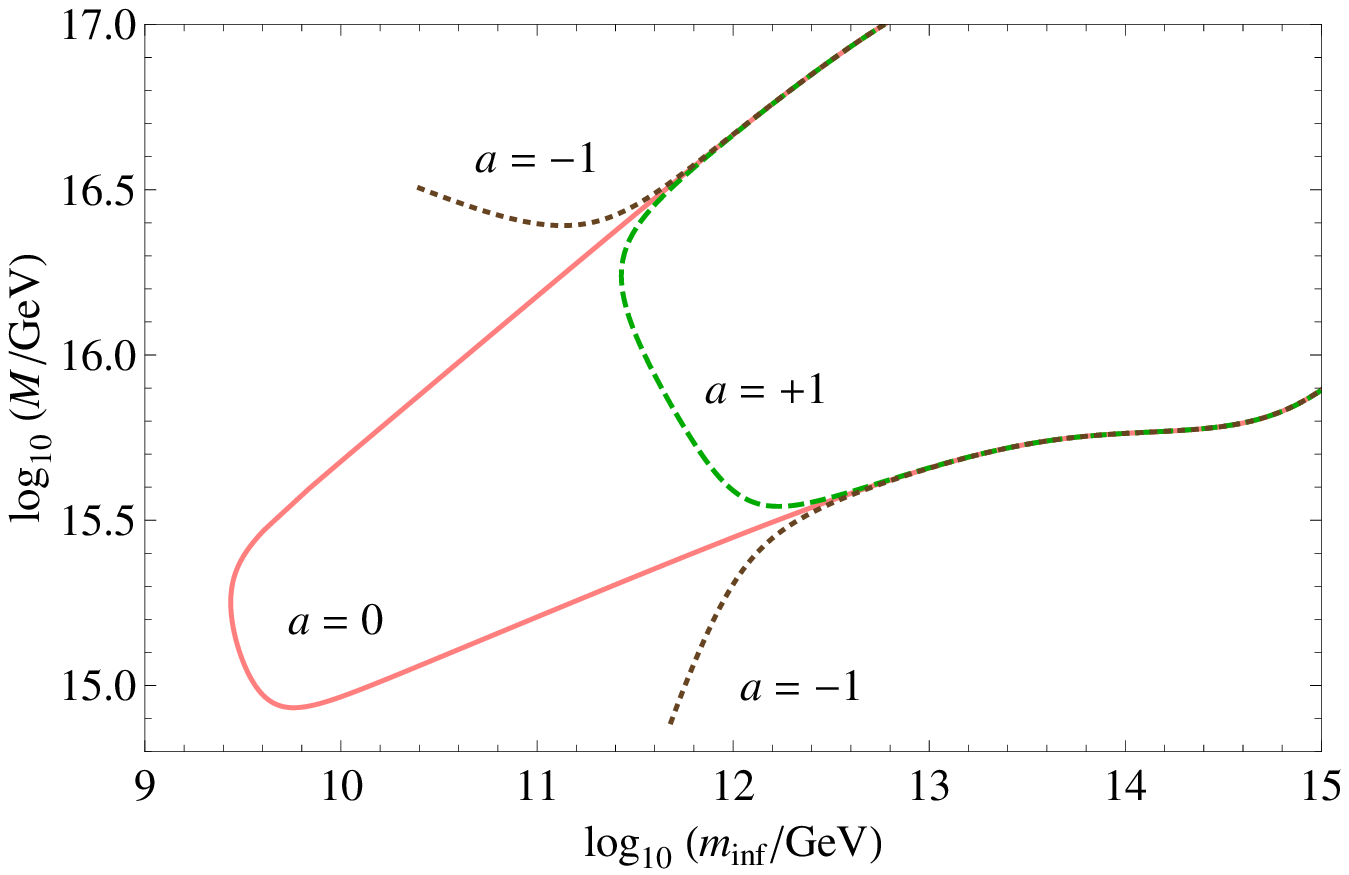}
\caption{$\log_{10}(T_r/\text{GeV})$ vs. $\kappa$ and $\log_{10}(M/\text{GeV})$ vs. $\log_{10}(m_{\text{inf}}/\text{GeV})$ for $m_{3/2} = 1$ TeV, $a = 0$, $a = -1$ and $a=+1$ (right panel),
with $a$ and $T_r$ given by Eqs.~(\ref{a}) and (\ref{trmin}), respectively.} \label{Trk}
\end{figure}

An important constraint on supersymmetric inflation models
arises from considering the reheat temperature $T_{r}$
after inflation, taking into account the gravitino problem which requires that
$T_{r}\lesssim10^6$--$10^{10}$ GeV \cite{Khlopov:1984pf}. This
constraint on $T_r$ depends on the SUSY breaking mechanism and the
gravitino mass $m_{3/2}$. For gravity mediated SUSY breaking models
with unstable gravitinos of mass $m_{3/2}\simeq0.1$--1~TeV,
$T_r\lesssim10^6$--$10^9$~GeV \cite{Kawasaki:1994af}, whereas
$T_r\lesssim10^{10}$~GeV for stable gravitinos \cite{Bolz:2000fu}.
The behavior of $T_r$ as a function of $\kappa$ is shown in Fig.~\ref{Trk}.  In the WMAP5 favored region (for $a=-1$), $T_r\sim 2\times 10^8$~GeV.  For $a=0$, the reheat temperature obtains a lower bound $T_r\sim 3\times 10^6$~GeV.
In addition, the temperature scale of reheating should always be lower than the mass of the inflaton, $m_{\rm inf}$.

Besides the upper bound placed on $T_r$ by thermal production, there are also constraints arising from non-thermal gravitino production via the direct decay of the inflaton.  While these constraints can be rather severe for SUSY hybrid inflation \cite{Kawasaki:2006gs}, gravitino production depends on the SUSY breaking sector and these models are still viable.  As displayed in Fig.~\ref{Trk}, significantly lower values of $m_{\rm inf}$ can be
obtained with $a < 1$. This extends the allowed range of parameters where the gravitino constraint can be evaded.  With minimal K\"ahler potential, it can be shown that even the most stringent constraint can be satisfied if $a$ is sufficiently small.

To summarize, we have investigated the important role played during hybrid inflation by one of the soft supersymmetry breaking terms generated in minimal supergravity. In the simplest example a scalar spectral index value of $n_s =0.963$, preferred by WMAP5, is realized if the overall sign of this term is negative. This case also permits one to estimate the inflaton mass to be of order $10^{9}-10^{12}$~GeV. It is amusing that this same soft term, linear in the inflaton field, also allows one to resolve the MSSM $\mu$ problem. Thus we conclude that supersymmetric hybrid inflation models with minimal K\"ahler potential are both realistic and in good agreement with the current observations. An important prediction of these models has to do with the tensor to scalar ratio $r$. We find that $r \leq 10^{-5}$ for $0.928 \leq n_s \leq 1.008$. 

\section*{Acknowledgments}
We thank Nefer {\c S}eno$\breve{\textrm{g}}$uz for valuable discussions. This work is
supported in part by the DOE under grant \# DE-FG02-91ER40626, by the University of Delaware competitive fellowship (M.R.), by NASA and the Delaware Space Grant Consortium under grant \#~NNG05GO92H (J.W.), and
by the Bartol Research Institute (M.R. and J.W.).

\end{document}